\documentclass[twocolumn,showpacs,showkeys,preprintnumbers,amsmath,amssymb]{revtex4}

\usepackage{graphicx}

\begin{document}


\title{Knee structure in high-energy inverse Compton scattering of CMB photons}
\author{Satoshi Nozawa}
 \email{snozawa@josai.ac.jp}
\affiliation{
Josai Junior College, 1-1 Keyakidai, Sakado-shi, Saitama, 350-0295,
Japan}

\author{Yasuharu Kohyama and Naoki Itoh}
\affiliation{
Department of Physics, Sophia University, 7-1 Kioi-cho, Chiyoda-ku,
Tokyo, 102-8554, Japan}

\date{\today}

\begin{abstract}
  We study the inverse Compton scattering of the CMB photons off nonthermal high-energy electrons.  In the previous study, assuming the power-law distribution for electrons, we derived the analytic expression for the spectral intensity function $I(\omega)$ in the Thomson approximation, which was applicable up to the photon energies of $\omega <$ O(GeV).  In the present paper, we extend the previous work to higher photon energies of $\omega >$ O(GeV) by taking into account the terms dropped in the Thomson approximation, i.e., the Klein-Nishina formula.  The analytic expression for $I(\omega)$ is derived with the Klein-Nishina formula.  It is shown that $I(\omega)$ has a ``knee" structure at $\omega =$ O(PeV).  The knee, if exists, should be accessible with gamma-ray observatories such as Fermi-LAT.  We propose simple analytical formulae for $I(\omega)$ which are applicable to wide photon energies from Thomson region to extreme Klein-Nishina region.
\end{abstract}



\maketitle

\section{Introduction}

  The inverse Compton scattering is one of the most fundamental reactions which have variety of applications to astrophysics and cosmology.  They are, for example, the Sunyaev-Zeldovich (SZ) effects\cite{suny72} for clusters of galaxies (CG), cosmic-ray emission from radio galaxies\cite{blun06} and clusters of galaxies\cite{sara99}, and radio to gamma-ray emission from supernova remnants\cite{bari99, laze04}.  Therefore, theoretical studies on the inverse Compton scattering have been done quite extensively for the last forty years, starting from the works by Jones\cite{jone68}, and Blumenthal and Gould\cite{blum70} to the recent works, for example, by Fargion\cite{farg97}, Colafrancesco\cite{cola08, cola09}, and Petruk\cite{petr09}.

  In particular, remarkable progress has been made in theoretical studies for the SZ effects for CG.  Wright\cite{wrig79} and Rephaeli\cite{reph95} calculated the photon frequency redistribution function in the electron rest frame, which is called as the radiative transfer method.  On the other hand, Challinor and Lasenby\cite{chal98} and Itoh, Kohyama, and Nozawa\cite{itoh98} solved the relativistically covariant Boltzmann collisional equation for the photon distribution function, which is called the covariant formalism.  Very recently, however, Nozawa and Kohyama\cite{noza09a} showed that the two formalisms were mathematically equivalent in the approximation of the Thomson limit.

  On the other hand, Nozawa, Kohyama and Itoh\cite{noza10a,noza10b} (denoted NKI hereafter) extended the formalism obtained by \cite{noza09a} to the case of high-energy electrons in the Thomson approximation.  Assuming the power-law distribution for electrons, the analytic expression was derived for the spectral intensity function $I(\omega)$, where $\omega$ is the photon energy.  This extension was particularly interesting for the analysis of X-ray and gamma-ray emissions, for example, from radio galaxies\cite{blun06} and supernova remnants\cite{bari99, laze04}, where the inverse Compton scattering of the CMB photons off nonthermal high-energy electrons plays an essential role.

  In the present paper, we extend the formalism obtained by NKI to higher photon energies beyond the Thomson approximation by calculating the Klein-Nishina formula.  The present extension will be particularly important to study the inverse Compton scattering of the CMB photons for the photon energies $\omega >$ O(GeV).  We will derive the analytical expression of $I(\omega)$ for the case of the power-law electron distribution with the Klein-Nishina formula.  It will be demonstrated that $I(\omega)$ has a ``knee" structure at $\omega =$ O(PeV).  The knee, if exists, should be accessible with gamma-ray observatories such as {\it Fermi} Gamma-ray Space Telescope's Large Area Telescope (Fermi-LAT)\cite{atwo09}.  We propose simple analytical formulae for $I(\omega)$ which are applicable to the entire photon energies.

  The present paper is organized as follows:  In Sec.~II, we summarize the results of the inverse Compton scattering in the Thomson approximation, which was obtained by NKI.  We calculate the rate equation and show the analytic expression for the spectral intensity function $dI(\omega)/d\tau$ in the Thomson approximation.  In Sec.~III, we extend the formalism of Sec.~II with the Klein-Nishina formula, and derive the analytical formula which is applicable to higher photon energies of $\omega >$ O(GeV).  In Sec.~IV, we discuss various other cases for the boundary parameters of the electron distribution function which are not studied in Sec.~III.  Finally, concluding remarks are given in Sec.~V.

\section{High-energy Inverse Compton scattering in Thomson Approximation}

\subsection{Rate equations}

  In the present section, we summarize the results obtained by NKI\cite{noza10a} in order to make the present paper more self-contained.  The rate equation for the photon distribution $n(\omega)$ and the spectral intensity function $I(\omega)$ were derived under the following Thomson approximation:
\begin{eqnarray}
&&\hspace{-10mm}
\gamma \frac{\omega}{m} \ll 1  \, ,
\label{eq2a-1}
\end{eqnarray}
where $\omega$ is the photon energy, $\gamma$ is the Lorentz factor, and $m$ is the rest mass of the electron.  Throughout this paper, we use the natural unit $\hbar = c = 1$, unless otherwise stated explicitly.  They are given as follows:
\begin{eqnarray}
&&\hspace{-10mm}
\frac{\partial n(\omega)}{\partial \tau}
 = \int_{-\infty}^{\infty}ds P_{1, \rm TA}(s)
\left[n(e^sx_{\rm T})- n(x_{\rm T})\right] \, ,
\label{eq2a-2}   \\
&&\hspace{-10mm}
\frac{\partial I(\omega)}{\partial \tau}
 = \int_{-\infty}^{\infty}ds
{P}_{1, \rm TA}(s) \left[I(e^{-s}x_{\rm T})- I(x_{\rm T})\right]  \, ,
\label{eq2a-3}  \\
&&\hspace{-10mm}
P_{1, \rm TA}(s) = \int_{\gamma_{*}}^{\infty}d\gamma \, p_e(\gamma) \, P_{\rm TA}(s,\gamma)  \, ,
\label{eq2a-4}  \\
&&\hspace{+12mm}
\tau  =  n_e\sigma_{\rm T} t \, , 
\label{eq2a-5}  \\
&&\hspace{+10mm}
x_{\rm T} =  \frac{\omega}{k_{\rm B} T_{\rm CMB}} \, .
\label{eq2a-6}
\end{eqnarray}
The subscript TA denotes that the Thomson approximation is used in the calculation.  In above equations, $I(\omega)=I_{0}x_{\rm T}^{3}n(\omega)$, $I_{0}=(k_{B}T_{\rm CMB})^{3}/2\pi^{2}$, $n_{e}$ is the electron number density, $\sigma_{\rm T}$ is the Thomson scattering cross section, and $P_{1, \rm TA}(s)$ is the probability distribution function for the photon of a frequency shift $s$ defined by $e^{s}=x_{\rm T}^{\prime}/x_{\rm T}$.

  In the present paper, we are interested in the inverse Compton scattering of the CMB photons by high-energy electrons, where $\gamma \gg 1$ is assumed.  In this approximation, the minimum value $\gamma_{*}$ of the $\gamma$-integration is approximated by
\begin{eqnarray}
&&\hspace{-10mm}
\gamma_{*} = \frac{1}{2} e^{\mid s \mid/2}  \, .
\label{eq2a-7}
\end{eqnarray}
Then, we have\cite{noza10a} the analytic expression for $P_{\rm TA}(s,\gamma)$ as follows:
\begin{eqnarray}
&&\hspace{-10mm}
P_{\rm TA}(s,\gamma) = \frac{3e^{3s/2}}{32\gamma^{2}} \left[
- \frac{1}{\gamma^{4}} e^{3 \mid s \mid /2} + \frac{2}{\gamma^{2}} e^{\mid s \mid /2} \right.
\nonumber  \\
&&\hspace{+17mm}
\left.
 + 8 e^{-\mid s \mid /2} + \frac{4 e^{\mid s \mid /2}}{\gamma^{2}} \ln \frac{ e^{ \mid s \mid}}{4 \gamma^{2}} \right]   \, .
\label{eq2a-8}
\end{eqnarray}

  In Eq.~(\ref{eq2a-4}), $p_{e}(\gamma)$ is the electron distribution function, which is normalized by $\int d \gamma p_{e}(\gamma) =1$.  High-energy electrons in the supernova remnants and active galactic nuclei, for example, are most likely nonthermal.  Therefore, it is standard to describe the nonthermal distribution in terms of the power-law distribution function of three parameters:
\begin{eqnarray}
&&\hspace{-10mm}
p_{e}(\gamma) = \left\{
\begin{array}{ll}
 N_{\gamma} \, \gamma^{-\sigma} \, , &  \, \, \, \gamma_{\rm min} \leq \gamma \leq \gamma_{\rm max} \\
   0  \, , &  \, \, \, {\rm elsewhere}
\end{array}
\right.  \, ,
\label{eq2a-9}
\end{eqnarray}
where $N_{\gamma}$ is the normalization constant, $\sigma$ is the power index parameter, $\gamma_{\rm min}$ and $\gamma_{\rm max}$ are parameters of minimum and maximum values for $\gamma$, respectively.

\subsection{Spectral intensity function}

  In NKI, the rate equations of Eqs.~(\ref{eq2a-2}) and (\ref{eq2a-3}) were solved for the Planckian initial CMB photons with the power-law electron distribution of Eq.~(\ref{eq2a-9}).  For the inverse Compton scattering of the CMB photons by high-energy electrons, we are interested in high-energy spectrum such as X-rays ($\sim$ keV) and gamma-rays ($\sim$ MeV).  Therefore, one can safely assume $x_{\rm T} \gg 1$ for scattered photons, where the second terms in Eqs.~(\ref{eq2a-2}) and (\ref{eq2a-3}) can be neglected.  For the $\gamma$-parameters, we assume the following condition:
\begin{eqnarray}
&&\hspace{-10mm}
1 \ll \gamma_{\rm min} \ll \gamma_{\rm max}  \, .
\label{eq2b-1}
\end{eqnarray}
Under these assumptions, Eqs.~(\ref{eq2a-2}) and (\ref{eq2a-3}) were solved analytically.  One obtained as follows:
\begin{eqnarray}
&&\hspace{-5mm}
\frac{d I_{\rm TA}(\omega)}{d \tau}= I_0 (\sigma - 1)
\left[ 3 X_{\rm T} \int_{X_{\rm T}}^{\infty} dt \frac{t}{e^t-1}
\left\{ \frac{1}{\sigma+1} \right. \right.
\nonumber  \\
&&\hspace{0mm}
\left. + \frac{1}{\sigma+3} \left( \frac{\sigma-1}{\sigma+3} + 2\ln\frac{X_{\rm T}}{t} \right) \frac{X_{\rm T}}{t} - \frac{2}{\sigma+5} \left( \frac{X_{\rm T}}{t} \right)^{2} \right\}
\nonumber   \\
&&\hspace{0mm}
+  \frac{6(\sigma^2+4\sigma+11)}{(\sigma+1)(\sigma+3)^2(\sigma+5)}
X_{\rm T}^{-(\sigma-1)/2}
\nonumber  \\
&&\hspace{30mm}
 \times \int_{0}^{X_{\rm T}}dt\frac{t^{(\sigma+3)/2}}{e^t-1} \Bigg]
\label{eq2b-2}  \, ,  \\
&&\hspace{0mm}
X_{\rm T} = \frac{x_{\rm T}}{4 \gamma_{min}^{2}} 
\label{eq2b-3}  \, .
\end{eqnarray}
Note that $N_{\gamma}=(\sigma-1)\gamma_{\rm min}^{\sigma-1}$ was used under the condition of Eq.~(\ref{eq2b-1}).

  For $X_{\rm T} \gg 1$, Eq.~(\ref{eq2b-2}) was further simplified as follows:
\begin{eqnarray}
&&\hspace{-10mm}
\frac{d I_{\rm TA}(\omega)}{d \tau} = I_0
\frac{6(\sigma-1)(\sigma^2+4\sigma+11)}{(\sigma+1)(\sigma+3)^2(\sigma+5)}
\nonumber \\
&&\hspace{3mm}
\times \Gamma\left(\frac{\sigma+5}{2}\right)
\zeta\left(\frac{\sigma+5}{2}\right) X_{\rm T}^{-(\sigma-1)/2}
\label{eq2b-4}  \, ,  \\
&&\hspace{-5mm}
\zeta(z) = \frac{1}{\Gamma(z)} \int_{0}^{\infty} dt \frac{t^{z-1}}{e^{t}-1}
\label{eq2b-5}  \, ,
\end{eqnarray}
where $\zeta(z)$ is the Riemann's zeta function.  In Fig.~3 of NKI, $dI_{\rm TA}(\omega)/d\tau$ of Eq.~(\ref{eq2b-2}) were plotted as a function of $X_{\rm T}$ for typical $\sigma$-values.  It was shown that Eq.~(\ref{eq2b-4}) was a good approximation of Eq.~(\ref{eq2b-2}) for $X_{\rm T} > 10$.

\section{High-energy Inverse Compton scattering with Klein-Nishina formula}

\subsection{Rate equations}

  In the present section, we extend the formalism developed in NKI\cite{noza10a} to higher photon energies where the Thomson approximation breaks down.  In order to accomplish the task, one needs to relax the condition of Eq.~(\ref{eq2a-1}).  Instead, one needs to keep the full (so called Klein-Nishina) amplitude for the Compton scattering process.
  
  Very recently, Poutanen and Vurm\cite{pout10} made an extensive study on the Compton scattering process with the general Klein-Nishina cross section.  The formalism can be applicable to higher energies.  On the other hand, it was shown by Nozawa, Kohyama and Itoh\cite{noza10c} that their formalism was equivalent to the NKI formalism in the Thomson approximation.  In the present paper, therefore, we start with the formalism of \cite{pout10} and derive the analytic expression for the spectral intensity function.

  The source function is defined by Eq.~(174) of \cite{pout10} and the rate equations for the photon number density and the spectral intensity function are given as follows:
\begin{eqnarray}
&&\hspace{-10mm}
\frac{\partial n(\omega)}{\partial \tau}
 = \int_{-\infty}^{\infty}ds P_{1, \rm KN}(s)
\left[n(e^sx_{\rm T})- n(x_{\rm T})\right] \, ,
\label{eq3a-1}   \\
&&\hspace{-10mm}
\frac{\partial I(\omega)}{\partial \tau}
 = \int_{-\infty}^{\infty}ds
{P}_{1, \rm KN}(s) \left[I(e^{-s}x_{\rm T})- I(x_{\rm T})\right]  \, ,
\label{eq3a-2}  \\
&&\hspace{-10mm}
P_{1, \rm KN}(s) = \int_{\gamma_{*}}^{\infty}d\gamma \, p_e(\gamma) \, P_{\rm KN}(s, \gamma)  \, ,
\label{eq3a-3}  \\
&&\hspace{-10mm}
P_{\rm KN}(s, \gamma) = \frac{3}{16\beta\gamma^2} x e^{2s} \int_{\mu_{\rm{min}}}^{\mu_{\rm{max}}} d\mu R_0  \, ,
\label{eq3a-4}  \\
&&\hspace{+3mm}
x = \frac{\omega}{mc^{2}}  \, .
\label{eq3a-5}
\end{eqnarray}
The subscript KN denotes that the Klein-Nishina cross section is used in the calculation.  In Eq.~(\ref{eq3a-3}), $\gamma_{*}$ is the minimum value of $\gamma$.  In Eq.~(\ref{eq3a-4}), $\mu_{\rm min}$ and $\mu_{\rm max}$ are minimum and maximum values of the cosine of the scattering angle, and the function $R_{0}$ is given by Eq.~(138) of \cite{pout10}.

  Let us consider the high-energy ($x \gg 1$) scattering of the CMB photons ($x^{\prime} \ll 1$) by extremely high-energy electrons ($\gamma \gg 1$, therefore $\beta \approx 1$).  It should be emphasized here that all conditions made in Sec.~II are retained except for Eq.~(\ref{eq2a-1}).  In this approximation, one finds
\begin{eqnarray}
&&\hspace{-10mm}
\gamma_{*} = \frac{1}{2} \left( x + x \sqrt{1 + \frac{1}{x x^{\prime}}} \right)  \, ,
\label{eq3a-6}  \\
&&\hspace{-10mm}
\mu_{\rm min} = -1  \, ,
\label{eq3a-7}  \\
&&\hspace{-10mm}
\mu_{\rm max} = 1 - \alpha \, ,
\label{eq3a-8}  \\
&&\hspace{-10mm}
\alpha \equiv \frac{e^{\mid s \mid}}{2\gamma (\gamma - x)}  \, .
\label{eq3a-9}
\end{eqnarray}
Then, Eq.~(\ref{eq3a-4}) can be approximated as follows:
\begin{eqnarray}
&&\hspace{-10mm}
P_{\rm KN}(s, \gamma) = \frac{3}{16\gamma^2} \Bigg[ 2e^{2s} I_{1} + x e^{2s} \left\{ I_{2}(\gamma - x) - I_{2}(\gamma) \right\}
\nonumber  \\
&&\hspace{+20mm}
- \frac{2e^{s}}{x} \left\{ I_{3}(\gamma - x) - I_{3}(\gamma) \right\}
\nonumber  \\
&&\hspace{+20mm}
- \frac{2}{x^{3}} \left\{ I_{4}(\gamma - x) - I_{4}(\gamma) \right\}
\nonumber  \\
&&\hspace{+20mm}
+ \frac{1}{x^{2}} \left\{ (\gamma - x) I_{5}(\gamma - x) - \gamma I_{5}(\gamma) \right\}
\nonumber  \\
&&\hspace{+20mm}
\left. - \frac{e^{s}}{x} \left\{ I_{6}(\gamma - x) - I_{6}(\gamma) \right\}
 \right]  \, ,
\label{eq3a-10}
\end{eqnarray}
where $I_{1}, I_{2}(\xi), \dots, I_{6}(\xi)$ are the integrals defined by
\begin{eqnarray}
&&\hspace{-10mm}
I_1 =
\int_{-1}^{\mu_{\rm{max}}}d\mu
\frac{1}{(1+e^{2s}-2e^s\mu)^{1/2}}  \, ,
\label{eq3a-11} \\
&&\hspace{-10mm}
I_2(\xi) =
\int_{-1}^{\mu_{\rm{max}}}d\mu
\frac{1}{(\xi^2+r)^{1/2}}  \, ,
\label{eq3a-12} \\
&&\hspace{-10mm}
I_3(\xi) =
\int_{-1}^{\mu_{\rm{max}}}d\mu
\frac{1}{(1-\mu)(\xi^2+r)^{1/2}}  \, ,
\label{eq3a-13} \\
&&\hspace{-10mm}
I_4(\xi) =
3\int_{-1}^{\mu_{\rm{max}}}d\mu
\frac{1}{(1-\mu)^2(\xi^2+r)^{1/2}}  \, ,
\label{eq3a-14} \\
&&\hspace{-10mm}
I_5(\xi) =
\int_{-1}^{\mu_{\rm{max}}}d\mu
\frac{1}{(1-\mu)^2(\xi^2+r)^{3/2}}  \, ,
\label{eq3a-15}  \\
&&\hspace{-10mm}
I_6(\xi) =
\int_{-1}^{\mu_{\rm{max}}}d\mu
\frac{1}{(1-\mu)(\xi^2+r)^{3/2}}  \, ,
\label{eq3a-16}  \\
\nonumber  \\
&&\hspace{+10mm}
r = \frac{1 + \mu}{1 - \mu}  \, .
\label{eq3a-17}
\end{eqnarray}
Their elementary integrals can be performed.  One has as follows:
\begin{eqnarray}
&&\hspace{-10mm}
I_1 = e^{-s}\left[1+e^s - \left\{1 - 2(1-\alpha)e^s + e^{2s} \right\}^{1/2} \right]  \, ,
\label{eq3a-18}
\end{eqnarray}
\begin{eqnarray}
&&\hspace{-10mm}
I_2(\xi) = - \frac{2}{\xi^2-1} \left( \frac{\alpha}{2} \sqrt{\xi^2+\delta} - \xi \right)
\nonumber  \\
&&\hspace{-5mm}
+ \frac{1}{(\xi^2-1)^{3/2}} \left[ \ln \frac{\alpha}{2} - \ln \left\{ 1 - \frac{1}{2\xi^2} + \sqrt{1 - \frac{1}{\xi^2}} \right\} \right.
\nonumber  \\
&&\hspace{-5mm}
\left. + \ln \left\{ 1 + \frac{1}{\xi^2} \left( \frac{1}{\alpha} - 1 \right) + \sqrt{1 + \frac{\delta}{\xi^2}} \sqrt{1 - \frac{1}{\xi^2}} \right\}  \right]    \, ,
\label{eq3a-19}
\end{eqnarray}
\begin{eqnarray}
&&\hspace{-10mm}
I_3(\xi) = - \frac{1}{\sqrt{\xi^2-1}} \left[ \ln \frac{\alpha}{2} - \ln \left\{ 1 - \frac{1}{2\xi^2} + \sqrt{1 - \frac{1}{\xi^2}} \right\} \right.
\nonumber  \\
&&\hspace{-5mm}
\left. + \ln \left\{ 1 + \frac{1}{\xi^2} \left( \frac{1}{\alpha} - 1 \right) + \sqrt{1 + \frac{\delta}{\xi^2}} \sqrt{1 - \frac{1}{\xi^2}} \right\}  \right]    \, ,
\label{eq3a-20}
\end{eqnarray}
\begin{eqnarray}
&&\hspace{-10mm}
I_4(\xi) = \sqrt{\xi^2 + \delta} - \xi  \, ,
\label{eq3a-21}
\end{eqnarray}
\begin{eqnarray}
&&\hspace{-10mm}
I_5(\xi) = \xi - \frac{1}{\sqrt{\xi^2 + \delta}}  \, ,
\label{eq3a-22}
\end{eqnarray}
\begin{eqnarray}
&&\hspace{-10mm}
I_6(\xi) = \frac{2}{\xi} - \frac{\alpha}{\sqrt{\xi^2 + \delta}} - I_2(\xi)    \, ,
\label{eq3a-23}
\end{eqnarray}
where
\begin{eqnarray}
&&\hspace{-10mm}
\delta \equiv \frac{2}{\alpha} - 1  \, .
\label{eq3a-24}
\end{eqnarray}

  Under assumptions of high-energy scattering, namely, $s \ll -1$, $\xi \gg 1$, and $\alpha = O(1)$, Eqs.~(\ref{eq3a-18})--(\ref{eq3a-23}) are further simplified by keeping only leading-order terms.  One has
\begin{eqnarray}
&&\hspace{-10mm}
I_1 = 2 - \alpha + O(e^s)  \, ,
\label{eq3a-25}  \\
&&\hspace{-10mm}
I_2(\xi) = \frac{2}{\xi} \left( 1 - \frac{\alpha}{2} \right) + O(\xi^{-3})   \, ,
\label{eq3a-26}  \\
&&\hspace{-10mm}
I_3(\xi) = - \frac{1}{\xi} \ln \frac{\alpha}{2} + O(\xi^{-3})  \, ,
\label{eq3a-27}  \\
&&\hspace{-10mm}
I_4(\xi) = \frac{1}{\xi} \left( \frac{1}{\alpha} - \frac{1}{2} \right) + O(\xi^{-3})  \, ,
\label{eq3a-28}  \\
&&\hspace{-10mm}
I_5(\xi) = \frac{2}{\xi^3} \left( 1 - \frac{\alpha}{2} \right) + O(\xi^{-5})  \, ,
\label{eq3a-29}  \\
&&\hspace{-10mm}
I_6(\xi) = O(\xi^{-5})  \, .
\label{eq3a-30}
\end{eqnarray}

  Inserting Eqs.~(\ref{eq3a-25})--(\ref{eq3a-30}) into Eq.~(\ref{eq3a-10}), and performing a lengthy but straightforward calculation, one finally obtains the following analytic expression for  $P_{\rm KN}(s, \gamma)$:
\begin{eqnarray}
&&\hspace{-8mm}
P_{\rm KN}(s,\gamma) = \frac{3e^{3s/2}}{32\gamma^{2}} \left[
- \frac{1}{\gamma^{2}(\gamma - x)^{2}} e^{3 \mid s \mid /2}  \right.
\nonumber  \\
&&\hspace{+8mm}
 + \frac{2}{\gamma(\gamma - x)} e^{\mid s \mid /2}
 + 8 e^{-\mid s \mid /2} + \frac{4 e^{\mid s \mid /2}}{\gamma(\gamma - x)} \ln \frac{\alpha}{2}
\nonumber  \\
&&\hspace{+8mm}
\left.  +  \frac{4x^2}{\gamma(\gamma - x)} \left(1 - \frac{\alpha}{2} \right) e^{ -\mid s \mid/2} \right]   \, ,
\label{eq3a-31}
\end{eqnarray}
where $\alpha$ is given by Eq.~(\ref{eq3a-9}).  It should be remarked that Eq.~(\ref{eq3a-31}) returns to Eq.~(\ref{eq2a-8}) by taking the Thomson limit $\gamma \gg x$ and $\alpha \rightarrow e^{\mid s \mid}/2\gamma^{2}$.  In deriving Eq.~(\ref{eq3a-31}), the formula
\begin{eqnarray}
&&\hspace{-10mm}
P_{\rm KN}(s,\gamma) = e^{3s} P_{\rm KN}(-s,\gamma)
\label{eq3a-32}
\end{eqnarray}
was also used in order to derive the expression for $s>0$.

\subsection{Spectral intensity function for $x > \gamma_{\rm min}$ region}

  In the present subsection, we now calculate the spectral intensity function with Eq.~(\ref{eq3a-31}).  In order to proceed the calculation, we use $p_{e}(\gamma)$ of Eqs.~(\ref{eq2a-9}) and (\ref{eq2b-1}) with an additional condition $x > \gamma_{\rm min}$.  For simplicity, we also put $\gamma_{\rm max} \rightarrow \infty$. (We study $x < \gamma_{\rm min}$ case and finite $\gamma_{\rm max}$ case in Secs.~IV-A and IV-B, respectively.)
  
  Let us first introduce new variables $u$, $t$ and $X$ as follows:
\begin{eqnarray}
&&\hspace{-10mm}
\gamma = x (1 + u)  \, ,
\label{eq3b-1}  \\
&&\hspace{-10mm}
t = e^{s} x_{\rm T}  \, ,
\label{eq3b-2}  \\
&&\hspace{-10mm}
X = \theta_{\rm CMB} x  \, ,
\label{eq3b-3}
\end{eqnarray}
where $\theta_{\rm CMB}=k_{\rm B}T_{\rm CMB}/mc^2$.  Then, we rewrite Eq.~(\ref{eq3a-31}) with these variables.  One obtains as follows:
\begin{eqnarray}
&&\hspace{-10mm}
\tilde{P}_{\rm KN}(t,u) = \frac{3 \theta_{\rm CMB}^2}{32 x^4 (1+u)^2} \left[ - \frac{1}{X^2} \frac{1}{u^2(1+u)^2}  \right.
\nonumber  \\
&&\hspace{+10mm}
+ \frac{2}{X} \frac{t}{u(1+u)} + 8t^2
\nonumber  \\
&&\hspace{+10mm}
- \frac{4}{X} \frac{t}{u(1+u)} \left\{ \ln 4Xt + \ln u(1+u)  \right\}
\nonumber  \\
&&\hspace{+10mm}
\left. + \frac{4t^2}{u(1+u)} - \frac{1}{X} \frac{t}{u^2(1+u)^2} \right]   \, .
\label{eq3b-4}
\end{eqnarray}

  Inserting Eqs.~(\ref{eq2a-9}) and (\ref{eq3b-4}) into Eq.~(\ref{eq3a-3}), and performing a straightforward calculation, one finally obtains as follows:
\begin{eqnarray}
&&\hspace{-5mm}
\frac{d I_{\rm KN}(\omega)}{d \tau}= \frac{3I_0}{32}(\sigma - 1) \left(\gamma_{\rm min} \theta_{\rm CMB}\right)^{\sigma-1} X^{-\sigma} Q_{\rm KN}(X, \sigma)  \, ,
\nonumber  \\
\label{eq3b-5}
\end{eqnarray}
\begin{eqnarray}
&&\hspace{-10mm}
Q_{\rm KN}(X, \sigma) = \int_{0}^{\infty} dt \frac{t}{e^t-1} \int_{u_{*}}^{\infty} du Q(X, t, u)  \, ,
\label{eq3b-6}
\end{eqnarray}
\begin{eqnarray}
&&\hspace{-8mm}
Q(X, t, u) = \frac{8}{(1+u)^{\sigma+2}}
 + \left(1 + \frac{1}{2Xt} - \frac{\ln 4Xt}{Xt}  \right) 
\nonumber  \\
&&\hspace{10mm}
\times \frac{1}{u(1+u)^{\sigma+3}} - \frac{4}{Xt} \frac{\ln [u(1+u)]}{u(1+u)^{\sigma+3}}
\nonumber  \\
&&\hspace{10mm}
- \frac{1}{Xt} \left( 1 + \frac{1}{Xt} \right) \frac{1}{u^{2}(1+u)^{\sigma+4}}   \, , 
\label{eq3b-7}
\end{eqnarray}
\begin{eqnarray}
&&\hspace{-10mm}
u_{*} = \frac{1}{2} \left( \sqrt{1 + \frac{1}{Xt}} - 1 \right)  \, ,
\label{eq3b-8}
\end{eqnarray}
where $u_{*}$ is the minimum value of the $u$-integration.  In deriving Eq.~(\ref{eq3b-6}), the Planckian distribution function was used for the initial CMB photons, and the second term in Eq.~(\ref{eq3a-2}) were neglected, because $x_{\rm T} \gg 1$ is valid.

  Equation (\ref{eq3b-6}) is one of the main results of the present paper.  The spectral intensity function $I_{\rm KN}(\omega)$ is expressed by the double integral form.  Therefore, the computation can be done extremely fast compared with the full numerical calculation of multi-dimensional integrals which involves numerical cancellations among various terms.

\subsection{Approximate forms for spectral intensity function}

The expression for $dI_{\rm KN}(\omega)/d\tau$ can be further simplified for $X \ll 1$ and $X \gg 1$ regions.  For example, $u_{*}$ can be approximated by
\begin{eqnarray}
&&\hspace{-10mm}
u_{*} = \left\{
\begin{array}{ll}
\displaystyle{\frac{1}{2 \sqrt{Xt}}}  &\quad  {\rm for} \, \, \, X \ll 1  \\
\\
\displaystyle{\frac{1}{4Xt}}  &\quad {\rm for} \, \, \, X \gg 1 
\end{array}
\right. \, ,
\label{eq3c-1}
\end{eqnarray}
where $t=O(1)$ is assumed.

\bigskip

\noindent
(i) $X \ll 1$ region

  First, for $X \ll 1$ region, the $u$-integration can be done by expanding Eq.~(\ref{eq3b-7}) in powers of 1/$u$.  Then, the $t$-integration can be also done, and one obtains the following analytic expression:
\begin{eqnarray}
&&\hspace{-5mm}
\frac{d I_{\rm KN}(\omega)}{d \tau} = I_{0}  \, X_{\rm T}^{-(\sigma-1)/2} \Biggr[
\nonumber  \\
&&\hspace{0mm}
\frac{6(\sigma - 1)(\sigma^2+4\sigma+11)}{(\sigma+1)(\sigma+3)^2(\sigma+5)}
\Gamma\left(\frac{\sigma+5}{2}\right)
\zeta\left(\frac{\sigma+5}{2}\right)
\nonumber  \\
&&\hspace{-2mm}
- \frac{6(\sigma - 1)(\sigma^2+6\sigma+16)}{(\sigma+4)^2(\sigma+6)}
\Gamma\left(\frac{\sigma+6}{2}\right)
\zeta\left(\frac{\sigma+6}{2}\right) \sqrt{X}
\nonumber  \\
&&\hspace{+20mm}
+ \, O(X) \, \Biggr]  \, .
\label{eq3c-2}
\end{eqnarray}
It is needless to mention that the condition $X_{\rm T} \gg 1$ is satisfied in this region.  The first term in Eq.~(\ref{eq3c-2}) coincides with Eq.~(\ref{eq2b-4}), and the second term is the first-order correction to the Thomson approximation.  It should be remarked that Eq.~(\ref{eq3c-2}) was originally derived by Blumenthal and Gould\cite{blum70} in their Eqs.~(2.76) and (2.77).

\bigskip

\noindent
(ii) $X \gg 1$ region

  On the other hand, for $X \gg 1$ region (so called extreme Klein-Nishina limit), the expression for $dI_{\rm KN}(\omega)/d\tau$ can be also simplified.  One obtains
\begin{eqnarray}
&&\hspace{-5mm}
\frac{d I_{\rm KN}(\omega)}{d \tau}= \frac{3I_0}{32}(\sigma - 1) \left(\gamma_{\rm min} \theta_{\rm CMB}\right)^{\sigma-1} X^{-\sigma} Q_{\rm KN}^{\rm ext}(X, \sigma)  \, ,
\nonumber  \\
\label{eq3c-3}
\end{eqnarray}
\begin{eqnarray}
&&\hspace{-10mm}
Q_{\rm KN}^{\rm ext}(X, \sigma) = 4 \zeta (2) \left[ \ln X + C(\sigma) + 1 - C_{E} - C_{\ell}  \right]  \, ,
\label{eq3c-4}
\end{eqnarray}
where $C_{E}$=0.5772 is the Euler's constant, and $C_{\ell}$ = 0.5770.  We show the explicit derivation of Eq.~(\ref{eq3c-4}) in Appendix A.  Note that the original expression was derived by Blumenthal and Gould\cite{blum70} in their Eq.~(2.88).  One can rewrite Eq.~(2.85) in \cite{blum70} as
\begin{eqnarray}
&&\hspace{-10mm}
C_{\rm BG}(\sigma) = \int_{0}^{\infty} \frac{du}{u} \left[ \frac{1}{(1+u)^{\sigma+3}} -  \frac{1}{1+u} \right]
\nonumber  \\
&&\hspace{10mm}
 + \frac{2}{\sigma+1} -1 +  2 \ln 2  \, .
\label{eq3c-5}
\end{eqnarray}
On the other hand, according to Eq.~(\ref{eqA-10}) we have
\begin{eqnarray}
&&\hspace{-10mm}
C(\sigma) = \int_{0}^{\infty} \frac{du}{u} \left[ \frac{1}{(1+u)^{\sigma+3}} -  \frac{1}{(1+u)^{n}} \right]
\nonumber  \\
&&\hspace{10mm}
 + \frac{2}{\sigma+1} -1  + 2 \ln 2 - \sum_{i=1}^{n-1} \frac{1}{i} \, .
\label{eq3c-6}
\end{eqnarray}
It is clear that $C_{BG}(\sigma)$ corresponds to $n=1$ of Eq.~(\ref{eq3c-6}).  In the present calculation, we choose the integer $n \approx \sigma + 3$ in order to minimize the integration in  Eq.~(\ref{eq3c-6}).

  It is important to note that Eq.~(\ref{eq3c-6}) is further simplified for $\sigma$ = integer cases.  One has
\begin{eqnarray}
&&\hspace{-10mm}
C(\sigma) = \frac{2}{\sigma+1} -1 + 2 \ln 2 - \sum_{i=1}^{\sigma+2} \frac{1}{i}  \, ,
\label{eq3c-7}
\end{eqnarray}
because the integrand in Eq.~(\ref{eq3c-6}) vanishes by choosing $n=\sigma+3$.  The analytic expression is also obtained for $\sigma$ = half-integer cases as follows:
\begin{eqnarray}
&&\hspace{-10mm}
C(\sigma) = \frac{2}{\sigma+1} -1 + 4 \ln 2 - 2\sum_{i=1}^{[\sigma]+3} \frac{1}{2i-1}  \, ,
\label{eq3c-8}
\end{eqnarray}
where $[\sigma]$ denotes the integer part of $\sigma$.  The explicit derivation is shown in Appendix B.  Similarly, it is shown that $C(\sigma)$ is written in analytic forms for $\sigma$ = arbitrary fractional numbers.  In Appendix B, we show their explicit forms.  These analytic forms will be extremely convenient to use compared with the original integral form.

\subsection{Knee structure of spectral intensity function}

  In the present paper, we have derived the expression for the spectral intensity function $dI(\omega)/d\tau$ in the Klein-Nishina region, where the Thomson approximation breaks down.  Equations (\ref{eq2b-2}) and (\ref{eq3b-5}) will provide reliable theoretical data of the high-energy inverse Compton scattering from Thomson region to extreme Klein-Nishina region.  It is known from Eqs.~(\ref{eq3c-2}) and (\ref{eq3c-3}) that the slope of the spectrum changes as
\begin{eqnarray}
&&\hspace{-10mm}
\frac{dI(\omega)}{d\tau} \propto  \left\{
\begin{array}{ll}
\omega^{-(\sigma-1)/2}  &\quad  {\rm for} \, \, \, X \ll 1  \\
\\
\omega^{-\sigma}  &\quad {\rm for} \, \, \, X \gg 1 
\end{array}
\right. \, .
\label{eq3d-1}
\end{eqnarray}
The transition of the slope is expected to occur at $X \approx 1$, which corresponds to
\begin{eqnarray}
&&\hspace{-10mm}
\omega \approx \frac{m^{2}c^{4}}{k_{\rm B}T_{\rm CMB}} \approx 10^{15} \, {\rm eV}  \, .
\label{eq3d-2}
\end{eqnarray}
However, the transition can be checked only through performing the full calculation with Eq.~(\ref{eq3b-7}).

\begin{figure}
\begin{center}
\includegraphics[angle=0,width=0.50\textwidth]{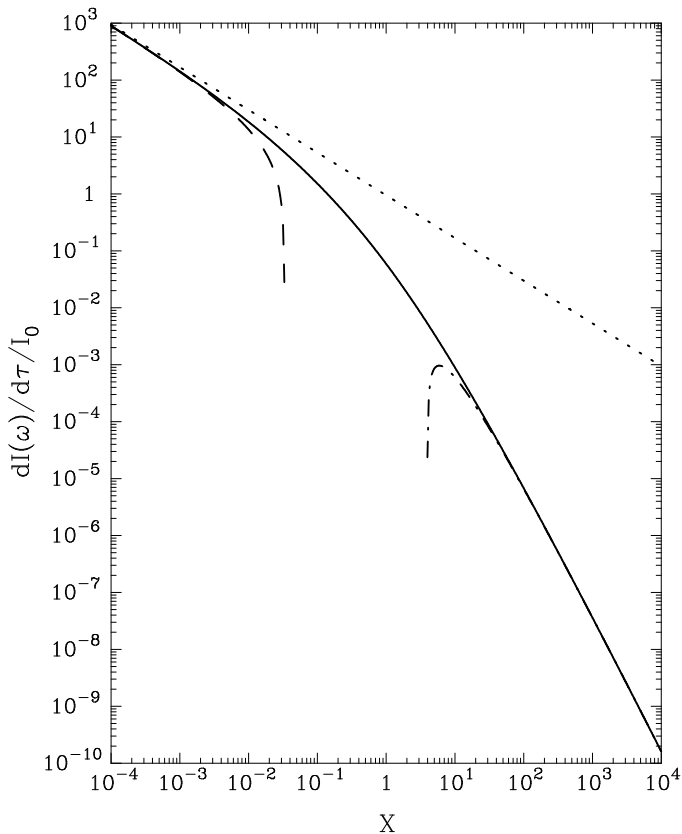}
\end{center}
\caption{Plotting of $dI(\omega)/d\tau$ as a function of $X$ for $\sigma=2.5$.  The solid curve is the full calculation of Eq.~(\ref{eq3b-5}) with Eq.~(\ref{eq3b-6}).  The dashed curve and dash-dotted curve correspond to the approximations of Eqs.~(\ref{eq3c-2}) and (\ref{eq3c-3}), respectively.  The dotted curve is the Thomson approximation of Eq.~(\ref{eq2b-4}).}
\end{figure}

  In Fig.~1, we have plotted $dI(\omega)/d\tau$ as a function of $X$ for a typical value $\sigma=2.5$.  The solid curve is the full calculation of Eq.~(\ref{eq3b-5}) with Eq.~(\ref{eq3b-6}).  The dashed curve and dash-dotted curve correspond to the approximations of Eqs.~(\ref{eq3c-2}) and (\ref{eq3c-3}), respectively.  The dotted curve is the Thomson approximation of Eq.~(\ref{eq2b-4}).  The full calculation clearly shows the knee structure at $X \approx 1$, which corresponds to $\omega \approx 10^{15}$ eV.  The knee, if exists, should be accessible with gamma-ray observatories such as Fermi-LAT\cite{atwo09}.

  On the other hand, the knee structure in the cosmic ray spectrum has been known since 1958\cite{kuli58}.  For the recent review, for example, see Bl\"{u}mer, Engel and H\"{o}randel\cite{blum09}.  The main sources of the cosmic ray are charged particles such as protons, He and heavier nuclei.  The observed energy spectrum shows a clear bending at around $10^{15}$ eV.  The energy spectrum follows a power law $dN/dE \propto E^{-a}$.  The power index increases from $a$=2.7 to $a$=3.1 at $E \approx 10^{15}$ eV, which gives the increment of $\Delta a$=0.4.  The origin of the knee structure, however, is yet unsettled.

  In the case of the inverse Compton scattering of the CMB photons, the increment of the power index is given by $\Delta a=(\sigma+1)/2=1.75$ for $\sigma=2.5$.  Although the position of the knee is same, the change of the slope is quite large compared with the case of the cosmic ray spectrum.

  In Fig.~2, we have shown a spectral intensity function $dI(\omega)/d\tau$ for a wide energy range from $\omega=10$ eV to $10^{21}$ eV.  A typical value $\gamma_{\rm min}$ = $10^{3}$ is used for an illustrative purpose.  The solid curve, dash-dotted curve and dashed curve correspond to $\sigma$ = 2.5, 3.5 and 4.5, respectively.  In the calculation, Eq.~(\ref{eq2b-2}) is used in the Thomson region ($\omega < 10^{9}$ eV), and Eq.~(\ref{eq3b-5}) with Eq.~(\ref{eq3b-6}) is used in the Klein-Nishina region ($\omega > 10^{9}$ eV).  The two curves coincide at the boundary region.  It should be remarkable that two theoretical formulae have the prediction power in the energy range of the twenty orders of magnitude.

\begin{figure}
\begin{center}
\includegraphics[angle=0,width=0.50\textwidth]{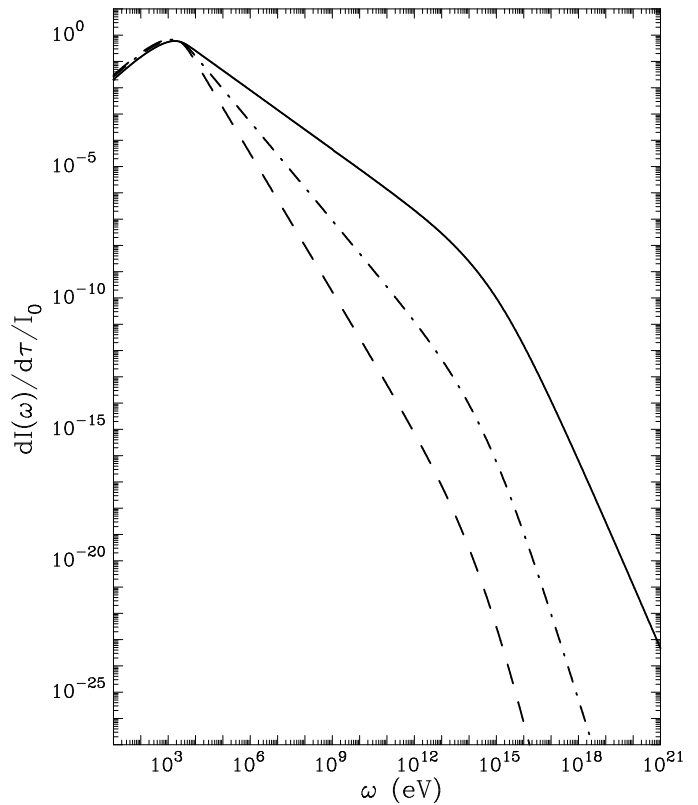}
\end{center}
\caption{Plotting of $dI(\omega)/d\tau$ as a function of $\omega$ for $\gamma_{\rm min}=10^{3}$.  The solid curve, dash-dotted curve and dashed curve correspond to $\sigma$ = 2.5, 3.5 and 4.5, respectively.  In the calculation, Eq.~(\ref{eq2b-2}) is used in the Thomson region ($\omega < 10^{9}$ eV), and Eq.~(\ref{eq3b-5}) with Eq.~(\ref{eq3b-6}) is used in the Klein-Nishina region ($\omega > 10^{9}$ eV).  The two curves coincide at the boundary region.}
\end{figure}

\section{Miscellaneous cases with Klein-Nishina formula}

\subsection{Spectral intensity function for $x < \gamma_{\rm min}$ region}

  In deriving the spectral intensity function with the Klein-Nishina formula in Sec.~III, we have restricted ourselves to the case where the photon energy $x$ satisfies the following condition (denoted Case I hereafter):
\begin{eqnarray}
&&\hspace{-10mm}
\gamma_{\rm min} < x \ll \gamma_{\rm max} \, , \gamma_{\rm max} \rightarrow \infty  \,\, \, \cdot \cdot \cdot \, \, \, {\rm Case \, \, I}  \, .
\label{eq4a-1}
\end{eqnarray}
In Case I, it is needless to mention that Eq.~(\ref{eq3b-5}) was not applicable to the photon energy region $x < \gamma_{\rm min}$.  For the $x < \gamma_{\rm min}$ region, however, Eq.~(\ref{eq3b-5}) was smoothly connected to Eq.~(\ref{eq2b-4}) of the Thomson approximation.

  In the present subsection, on the other hand, we derive the spectral intensity function with the Klein-Nishina formula for the $x < \gamma_{\rm min}$ region under the following condition (denoted Case II hereafter):
\begin{eqnarray}
&&\hspace{-10mm}
x < \gamma_{\rm min} \ll \gamma_{\rm max} \, , \gamma_{\rm max} \rightarrow \infty  \,\, \, \cdot \cdot \cdot \, \, \, {\rm Case \, \, II}  \, .
\label{eq4a-2}
\end{eqnarray}

  In order to accomplish the task, we first rewrite the rate equation of Eq.~(\ref{eq3a-2}) with the explicit lower ($\gamma_{\rm min}$) and upper ($\gamma_{\rm max}$) boundary parameters as follows:
\begin{eqnarray}
&&\hspace{-10mm}
\frac{\partial I(\omega)}{\partial \tau}
 = \int_{-s_{\rm max}}^{\infty}ds I(e^{-s}x_{\rm T})
\nonumber  \\
&&\hspace{5mm}
\times \int_{\rm max(\gamma_{*}, \gamma_{min})}^{\gamma_{\rm max}} d\gamma \, p_e(\gamma) \, P_{\rm KN}(s, \gamma)   \, ,
\label{eq4a-3}
\end{eqnarray}
\begin{eqnarray}
&&\hspace{-10mm}
s_{\rm max} = \ln [4\gamma_{\rm max} (\gamma_{\rm max} - x) ] \, ,
\label{eq4a-4}
\end{eqnarray}
where $\gamma_{*}$ is given by Eq.~(\ref{eq3a-6}), and we dropped the second term in Eq.~(\ref{eq3a-2}).  It is clear that Eq.~(\ref{eq4a-3}) returns to Eq.~(\ref{eq3a-2}) under the condition of Case I.

  Under the condition of Case II, Eq.~(\ref{eq4a-3}) becomes
\begin{eqnarray}
&&\hspace{-12mm}
\frac{\partial I(\omega)}{\partial \tau}
 = \int_{-\infty}^{-s_{\rm min}} ds I(e^{-s}x_{\rm T})
\int_{\gamma_{*}}^{\infty} d\gamma \, p_e(\gamma) \, P_{\rm KN}(s, \gamma)  
\nonumber  \\
&&\hspace{-2mm}
+ \int_{-s_{\rm min}}^{\infty} ds I(e^{-s}x_{\rm T})
\int_{\gamma_{\rm min}}^{\infty} d\gamma \, p_e(\gamma) \, P_{\rm KN}(s, \gamma)  \, ,
\label{eq4a-5}
\end{eqnarray}
\begin{eqnarray}
&&\hspace{-10mm}
s_{\rm min} = \ln [4\gamma_{\rm min} (\gamma_{\rm min} - x) ] \, .
\label{eq4a-6}
\end{eqnarray}
Repeating the same procedure done in Sec.~III, one finally obtains
\begin{eqnarray}
&&\hspace{-5mm}
\frac{d I^{\rm II}_{\rm KN}(\omega)}{d \tau}= \frac{3I_0}{32}(\sigma - 1) \left(\gamma_{\rm min} \theta_{\rm CMB}\right)^{\sigma-1} X^{-\sigma} Q^{\rm II}_{\rm KN}(X, \sigma)  \, ,
\nonumber  \\
\label{eq4a-7}
\end{eqnarray}
\begin{eqnarray}
&&\hspace{-10mm}
Q^{\rm II}_{\rm KN}(X, \sigma) = \int_{0}^{t_{\rm min}} dt \frac{t}{e^t-1} \int_{u_{*}}^{\infty} du Q(X, t, u)
\nonumber  \\
&&\hspace{8mm}
+ \int_{t_{\rm min}}^{\infty} dt \frac{t}{e^t-1} \int_{u_{\rm min}}^{\infty} du Q(X, t, u)  \, ,
\label{eq4a-8}
\end{eqnarray}
where $Q(X, t, u)$ is given by Eq.~(\ref{eq3b-7}), and the boundary values are expressed by
\begin{eqnarray}
&&\hspace{-10mm}
t_{\rm min} = \frac{1}{4 \Gamma_{\rm min} u_{\rm min}}  \, ,
\label{eq4a-9}  \\
&&\hspace{-10mm}
u_{\rm min} = \frac{\Gamma_{\rm min}}{X} - 1  \, ,
\label{eq4a-10}  \\
&&\hspace{-10mm}
\Gamma_{\rm min} = \theta_{\rm CMB} \gamma_{\rm min}  \, ,
\label{eq4a-11} 
\end{eqnarray}
and $u_{*}$ is given by Eq.~(\ref{eq3b-8}).  It is needless to mention that $Q^{\rm II}_{\rm KN}(X, \sigma)$ contains explicit $\gamma_{\rm min}$-dependences in contrast with $Q_{\rm KN}(X, \sigma)$.

  In the Thomson limit $x \ll \gamma_{\rm min}$, Eq.~(\ref{eq4a-7}) is further simplified.  Expanding $Q(X, t, u)$ in powers of $1/u$, the $u$-integration can be done analytically.  Then, it is straightforward to show that Eq.~(\ref{eq4a-7}) returns to Eq.~(\ref{eq2b-2}) as it should.

\subsection{Spectral intensity function for $x < \gamma_{\rm max}$ region}

  In the present subsection, we show the spectral intensity function with the Klein-Nishina formula for the case where the photon energy $x$ is not far away from the upper boundary, namely $x < \gamma_{\rm max}$ region under the following condition (denoted Case III hereafter):
\begin{eqnarray}
&&\hspace{-10mm}
\gamma_{\rm min} \ll x < \gamma_{\rm max}  \,\, \, \cdot \cdot \cdot \, \, \, {\rm Case \, \, III}  \, .
\label{eq4b-1}
\end{eqnarray}
  Under the condition of Case III, Eq.~(\ref{eq4a-3}) becomes
\begin{eqnarray}
&&\hspace{-8mm}
\frac{\partial I(\omega)}{\partial \tau}
 = \int_{-s_{\rm max}}^{\infty} ds I(e^{-s}x_{\rm T})
\int_{\gamma_{*}}^{\gamma_{\rm max}} d\gamma \, p_e(\gamma) \, P_{\rm KN}(s, \gamma)  \, .
\nonumber  \\
\label{eq4b-2}
\end{eqnarray}

Repeating the same procedure done in Sec.~III, one finally obtains
\begin{eqnarray}
&&\hspace{-5mm}
\frac{d I^{\rm III}_{\rm KN}(\omega)}{d \tau}= \frac{3I_0}{32}(\sigma - 1) \left(\gamma_{\rm min} \theta_{\rm CMB}\right)^{\sigma-1} X^{-\sigma} Q^{\rm III}_{\rm KN}(X, \sigma)  \, ,
\nonumber  \\
\label{eq4b-3}
\end{eqnarray}
\begin{eqnarray}
&&\hspace{-10mm}
Q^{\rm III}_{\rm KN}(X, \sigma) = \int_{t_{\rm max}}^{\infty} dt \frac{t}{e^t-1} \int_{u_{*}}^{u_{\rm max}} du Q(X, t, u)  \, ,
\label{eq4b-4}
\end{eqnarray}
where $Q(X, t, u)$ is given by Eq.~(\ref{eq3b-7}), and the boundary values are expressed by
\begin{eqnarray}
&&\hspace{-10mm}
t_{\rm max} = \frac{1}{4 \Gamma_{\rm max} u_{\rm max}}  \, ,
\label{eq4b-5}  \\
&&\hspace{-10mm}
u_{\rm max} = \frac{\Gamma_{\rm max}}{X} - 1  \, ,
\label{eq4b-6} \\
&&\hspace{-10mm}
\Gamma_{\rm max} = \theta_{\rm CMB} \gamma_{\rm max} \, ,
\label{eq4b-7} 
\end{eqnarray}
and $u_{*}$ is given by Eq.~(\ref{eq3b-8}).  Again, it is needless to mention that $Q^{\rm III}_{\rm KN}(X, \sigma)$ contains explicit $\gamma_{\rm max}$-dependences in contrast with $Q_{\rm KN}(X, \sigma)$.

  Finally, for the extreme Klein-Nishina limit under the condition of
\begin{eqnarray}
&&\hspace{-10mm}
X \gg 1, \, \Gamma_{\rm max} \gg 1  \, ,
\label{eq4b-8} 
\end{eqnarray}
Eq.~(\ref{eq4b-4}) is further simplified as follows:
\begin{eqnarray}
&&\hspace{-5mm}
Q_{\rm KN}^{\rm III,ext}(X, \sigma) = 4 \zeta (2) \left[ \ln X + C(X, \sigma) + 1 - C_{E} - C_{\ell}  \right]  \, ,
\nonumber  \\
\label{eq4b-10}
\end{eqnarray}
where
\begin{eqnarray}
&&\hspace{-8mm}
C(X, \sigma) = \int_{0}^{u_{\rm max}} \frac{du}{u} \left[ \frac{1}{(1+u)^{\sigma+3}} -  \frac{1}{(1+u)^{n}} \right]
\nonumber  \\
&&\hspace{5mm}
 + \frac{2}{\sigma+1} \left[ 1 - \left(\frac{X}{\Gamma_{\rm max}}\right)^{\sigma+1} \right] -1  + 2 \ln 2
\nonumber  \\
&&\hspace{5mm}
+ \ln \left(1 - \frac{X}{\Gamma_{\rm max}} \right) - \sum_{i=1}^{n-1} \frac{1}{i} \left[ 1 - \left( \frac{X}{\Gamma_{\rm max}} \right)^{i} \right] \, .
\nonumber  \\
\label{eq4b-11}
\end{eqnarray}
It is needless to mention that $C(X,\sigma)$ returns to $C(\sigma)$ of Eq.~(\ref{eq3c-6}) for $x \ll \gamma_{\rm max}$ (Case I).

\section{Concluding Remarks}

  In NKI\cite{noza10a,noza10b}, we have extended the formalism obtained by Nozawa and Kohyama\cite{noza09a} to the case of high-energy electrons in the Thomson approximation.  Assuming the power-law distribution for electrons, the analytic expression was derived for the spectral intensity function $dI(\omega)/d\tau$.

  In the present paper, we have extend the formalism developed in NKI to higher photon energy (so called Klein-Nishina) region where the Thomson approximation breaks down.  This work has been done on the basis of the recent work by Poutanen and Vurm\cite{pout10}, where the general Klein-Nishina cross section has been calculated for the study of the inverse Compton scattering process.

  We first derived the analytic form for the redistribution function $P_{\rm KN}(s,\gamma)$.  The obtained expression of $P_{\rm KN}(s,\gamma)$ is a natural extension of $P_{\rm TA}(s,\gamma)$ derived in the Thomson approximation.  Then, we have derived the expression for the spectral intensity function $dI_{\rm KN}(\omega)/d\tau$, where the function is expressed by the double integral form.  The form can be used in the entire Klein-Nishina energy region ($\omega > 10^{9}$ eV).  The function has an advantage that its computation can be done extremely fast compared with the full numerical calculation of multi-dimensional integrals which involve the numerical cancellations among various terms.

  We have also derived approximate expressions for $dI_{\rm KN}(\omega)/d\tau$, which agreed with Blumenthal and Gould\cite{blum70}.  The expression for $C(\sigma)$ derived in the present paper has an advantage that it can be written in analytic forms for $\sigma$ = arbitrary fractional numbers.

  We have studied the knee structure of $dI_{\rm KN}(\omega)/d\tau$.  The physics origin on the knee structure is clear.  The position of the knee is determined by $\omega=m^2c^4/k_{\rm B}T_{\rm CMB}$, which is the characteristic energy scale of the CMB photons and electrons.  The origin of the change of the slope in the spectrum is due to the change from the Thomson approximation to the extreme Klein-Nishina approximation.   It has been shown that the spectrum changes its slope at $\omega \approx 10^{15}$ eV from $a=(\sigma-1)/2$ to $a=\sigma$.  The position of the knee is same as the well-known cosmic ray spectrum.  However, the increment of the power index is $\Delta a=1.75$ for $\sigma=$2.5, which is quite large compared with $\Delta a=$0.4 of the cosmic ray spectrum.

  Finally, we have studied various cases of the boundary parameters ($\gamma_{\rm min}$ and $\gamma_{\rm max}$) of the electron distribution function, where the spectral intensity function depends explicitly on the $\gamma_{\rm min}$ and $\gamma_{\rm max}$ parameters.  For the extreme Klein-Nishina limit, we have shown an analytical expression for the spectral intensity function which takes into account for the finite size correction of the $\gamma_{\rm max}$ parameter.

\begin{acknowledgments}
This work is financially supported in part by the Grant-in-Aid of Japanese Ministry of Education, Culture, Sports, Science, and Technology under Contract No. 21540277.
\end{acknowledgments}

\appendix

\section{Derivation of $Q_{\rm KN}^{\rm ext}(X,\sigma)$}

  We derive the expression of $Q_{\rm KN}^{\rm ext}(X, \sigma)$ from Eq.~(\ref{eq3b-6}) by taking the limit $X \gg 1$.  According to Eq.~(\ref{eq3c-1}), one has $u_{*} \ll 1$.  The first term in Eq.~(\ref{eq3b-7}) can be integrated analytically.  One has
\begin{eqnarray}
&&\hspace{-10mm}
\int_{u_{*}}^{\infty} du \frac{1}{(1+u)^{\sigma+2}} \approx \frac{1}{\sigma+1}  \, .
\label{eqA-1}
\end{eqnarray}

  The second term is
\begin{eqnarray}
&&\hspace{-5mm}
\int_{u_{*}}^{\infty} du \frac{1}{u(1+u)^{\sigma+3}}
\approx \int_{u_{*}}^{\infty} du \frac{1}{u} \left[ \frac{1}{(1+u)^{\sigma+3}} - \frac{1}{(1+u)^{n}} \right]
\nonumber  \\
&&\hspace{28mm}
 + \ln 4Xt - \sum_{i=1}^{n-1} \frac{1}{i}  \, ,
\label{eqA-2}
\end{eqnarray}
where $n$ is a positive integer of the order of $\sigma+3$.  Equation (\ref{eqA-2}) can be shown as follows:  First, define $J_{n}$ by
\begin{eqnarray}
&&\hspace{-10mm}
J_{n} \equiv \int_{u_{*}}^{\infty} du \frac{1}{u(1+u)^{n}}  \,  .
\label{eqA-3}
\end{eqnarray}
Inserting the identity relation
\begin{eqnarray}
&&\hspace{-10mm}
\frac{1}{u(1+u)^{n}} = \frac{1}{u(1+u)^{n-1}} - \frac{1}{(1+u)^{n}}
\label{eqA-4}
\end{eqnarray}
into Eq.~(\ref{eqA-3}), and solving the recurrence relation for $J_{n}$, one obtains
\begin{eqnarray}
&&\hspace{-10mm}
J_{n} = J_{1} - \sum_{i=1}^{n-1} \frac{1}{i}  \,  ,
\label{eqA-5}  \\
&&\hspace{-10mm}
J_{1} = \int_{u_{*}}^{\infty} du \frac{1}{u(1+u)}
\nonumber  \\
&&\hspace{-3mm}
\approx \ln \frac{1}{u_{*}}  \approx \ln 4Xt  \,  .
\label{eqA-6}
\end{eqnarray}
One has
\begin{eqnarray}
&&\hspace{-10mm}
\int_{u_{*}}^{\infty} du \frac{1}{u(1+u)^{n}} \approx \ln 4Xt - \sum_{i=1}^{n-1} \frac{1}{i}  \, ,
\label{eqA-7}
\end{eqnarray}
Therefore, Eq.~(\ref{eqA-2}) is valid.

  The third term can be neglected, and the fourth term is
\begin{eqnarray}
&&\hspace{-10mm}
\int_{u_{*}}^{\infty} du \frac{1}{u^2(1+u)^{\sigma+4}}  \approx 4Xt  \,  .
\label{eqA-8}
\end{eqnarray}

  Inserting Eqs.~(\ref{eqA-1}), (\ref{eqA-2}) and (\ref{eqA-7}) into Eq.~(\ref{eq3b-6}), one obtains
\begin{eqnarray}
&&\hspace{-10mm}
Q_{\rm KN}^{\rm ext}(X, \sigma) = 4 \int_{0}^{\infty} dt \frac{t}{e^t-1} \left[ \ln t + \ln X + C(\sigma)  \right]  \, ,
\label{eqA-9}
\end{eqnarray}
\begin{eqnarray}
&&\hspace{-10mm}
C(\sigma) = \int_{0}^{\infty} \frac{du}{u} \left[ \frac{1}{(1+u)^{\sigma+3}} -  \frac{1}{(1+u)^{n}} \right]
\nonumber  \\
&&\hspace{3mm}
 + \frac{2}{\sigma+1} -1  + 2\ln 2 - \sum_{i=1}^{n-1} \frac{1}{i} \, ,
\label{eqA-10}
\end{eqnarray}
where we put $u_{*} \rightarrow 0$.  The $t$-integration of Eq.~(\ref{eqA-9}) can be done analytically as follows:
\begin{eqnarray}
&&\hspace{-5mm}
Q_{\rm KN}^{\rm ext}(X, \sigma) = 4 \zeta(2) \Gamma(2) \left[ \ln X + C(\sigma) + \frac{\zeta^{\prime}(2)}{\zeta(2)} + \frac{\Gamma^{\prime}(2)}{\Gamma(2)} \right]  \, .
\nonumber  \\
\label{eqA-11}
\end{eqnarray}
In deriving Eq.~(\ref{eqA-11}), we used
\begin{eqnarray}
&&\hspace{-10mm}
\int_{0}^{\infty} dt \frac{t^{s-1}}{e^t-1} = \zeta(s) \Gamma(s)  \, ,
\label{eqA-12}  \\
&&\hspace{-10mm}
\int_{0}^{\infty} dt \frac{t^{s-1} \ln t}{e^t-1} = \frac{d}{ds} \left[ \zeta(s) \Gamma(s) \right]
\nonumber  \\
&&\hspace{14mm}
= \zeta(s) \Gamma(s) \left[ \frac{\zeta^{\prime}(s)}{\zeta(s)} + \frac{\Gamma^{\prime}(s)}{\Gamma(s)} \right]  \, .
\label{eqA-13}
\end{eqnarray}
Inserting $\zeta^{\prime}(2)/\zeta(2)$ = $-6/\pi^{2}\sum_{k=2}^{\infty} \ln k/k^{2}$ = $-C_{\ell}$, $\Gamma^{\prime}(2)/\Gamma(2)$ = $1 - C_{E}$, and $\Gamma(2)=1$ into Eq.~(\ref{eqA-11}), one finally obtains
\begin{eqnarray}
&&\hspace{-5mm}
Q_{\rm KN}^{\rm ext}(X, \sigma) = 4 \zeta(2) \left[ \ln X + C(\sigma) + 1 - C_{E} - C_{\ell} \right]  \, ,
\nonumber  \\
\label{eqA-14}
\end{eqnarray}
where $C_{E} = 0.5772$ is the Euler's constant, and $C_{\ell} = 0.5700$.

\section{Analytic forms for $C(\sigma)$}

  We derive analytic forms of $C(\sigma)$ for specific $\sigma$-values.  For example, the analytic form was derived for $\sigma$ = integer cases in Eq.~(\ref{eq3c-7}).

  In order to derive the analytic expression for $\sigma$ = half-integer cases, let us introduce
\begin{eqnarray}
&&\hspace{-10mm}
J_{n}^{(\frac{1}{2})} \equiv \int_{u_{*}}^{\infty} du \frac{1}{u(1+u)^{n+\frac{1}{2}}}  \, ,
\label{eqB-1}
\end{eqnarray}
where $n \geq 0$ is an arbitrary integer, and $u_{*} \approx 1/4Xt \ll 1$.  Rewriting Eq.~(\ref{eqB-1}) with a new variable $v=1/(1+u)^{1/2}$, one obtains the following recurrence relation:
\begin{eqnarray}
&&\hspace{-10mm}
J_{n}^{(\frac{1}{2})} - J_{n-1}^{(\frac{1}{2})} = -\frac{2}{2n-1}  \, .
\label{eqB-2}
\end{eqnarray}
Equation (\ref{eqB-2}) can be solved, and one has
\begin{eqnarray}
&&\hspace{-10mm}
J_{n}^{(\frac{1}{2})} = J_{0}^{(\frac{1}{2})} -2 \sum_{i=1}^{n} \frac{1}{2i-1}  \, ,
\label{eqB-3}  \\
&&\hspace{-10mm}
J_{0}^{(\frac{1}{2})} \approx \ln \frac{4}{u_{*}} = \ln 16Xt  \, .
\label{eqB-4}
\end{eqnarray}
Therefore, one obtains
\begin{eqnarray}
&&\hspace{-10mm}
\int_{u_{*}}^{\infty} du \frac{1}{u(1+u)^{n+\frac{1}{2}}} = \ln 16Xt -2 \sum_{i=1}^{n} \frac{1}{2i-1}  \, .
\label{eqB-5}
\end{eqnarray}
Thus, we have
\begin{eqnarray}
&&\hspace{-10mm}
C(\sigma) = \int_{0}^{\infty} \frac{du}{u} \left[ \frac{1}{(1+u)^{\sigma+3}} -  \frac{1}{(1+u)^{n+\frac{1}{2}}} \right]
\nonumber  \\
&&\hspace{3mm}
 + \frac{2}{\sigma+1} -1  + 4 \ln 2  - 2 \sum_{i=1}^{n} \frac{1}{2i-1}  \, .
\label{eqB-6}
\end{eqnarray}
Finally, the analytic form for $\sigma$ = half-integers is obtained by inserting $n = [\sigma] + 3$ into Eq.~(\ref{eqB-6}) as follows:
\begin{eqnarray}
&&\hspace{-10mm}
C(\sigma) = \frac{2}{\sigma+1} -1 + 4 \ln 2 - 2\sum_{i=1}^{[\sigma]+3} \frac{1}{2i-1}  \, ,
\label{eqB-7}
\end{eqnarray}
where $[\sigma]$ denotes the integer part of $\sigma$.

  Similarly, one can derive the analytic forms for $\sigma=q/p$ (irreducible fractional numbers).  Let us introduce
\begin{eqnarray}
&&\hspace{-10mm}
J_{n}^{(\frac{q}{p})} \equiv \int_{u_{*}}^{\infty} du \frac{1}{u(1+u)^{n+\frac{q}{p}}}  \, ,
\label{eqB-8}
\end{eqnarray}
where $n$ is a natural number, and $p$ and $q$ are positive coprime integers.  Rewriting Eq.~(\ref{eqB-8}) with a new variable $v=1/(1+u)^{1/p}$, and solving the recurrence relation, one obtains
\begin{eqnarray}
&&\hspace{-10mm}
J_{n}^{(\frac{q}{p})} =  J_{0}^{(\frac{q}{p})}  - p \sum_{i=1}^{n} \frac{1}{pi - (p-q)}  \, ,
\label{eqB-9}  \\
&&\hspace{-10mm}
J_{0}^{(\frac{q}{p})} = p \int_{0}^{v_{*}} dv \frac{v^{q-1}}{1 - v^{p}}  \, ,
\label{eqB-10}
\end{eqnarray}
where $v_{*} \approx 1 - u_{*}/p$.  The elementary integral of Eq.~(\ref{eqB-10}) can be done analytically.

  We summarize the analytic expressions, for example, in the following cases:

\noindent
(i) for $\sigma=n$
\begin{eqnarray}
&&\hspace{-10mm}
C(\sigma) = \frac{2}{[\sigma]+1} -1 + 2 \ln 2 - \sum_{i=1}^{[\sigma]+2} \frac{1}{i}  \, ,
\label{eqB-11}
\end{eqnarray}
(ii) for $\sigma=n+\frac{1}{8}$
\begin{eqnarray}
&&\hspace{-10mm}
C(\sigma) = \frac{2}{[\sigma]+\frac{1}{8}+1} -1 + 6 \ln 2 + \frac{\pi}{2}
\nonumber  \\
&&\hspace{0mm}
 + \frac{\pi}{\sqrt{2}} + \sqrt{2} \ln (\sqrt{2}+1) - 8 \sum_{i=1}^{[\sigma]+3} \frac{1}{8i-7}  \, ,
\label{eqB-12}
\end{eqnarray}
(iii) for $\sigma=n+\frac{1}{4}$
\begin{eqnarray}
&&\hspace{-10mm}
C(\sigma) = \frac{2}{[\sigma]+\frac{1}{4}+1} -1 + 5 \ln 2 + \frac{\pi}{2}
\nonumber  \\
&&\hspace{0mm}
 - 4 \sum_{i=1}^{[\sigma]+3} \frac{1}{4i-3}  \, ,
\label{eqB-13}
\end{eqnarray}
(iv) for $\sigma=n+\frac{3}{8}$
\begin{eqnarray}
&&\hspace{-10mm}
C(\sigma) = \frac{2}{[\sigma]+\frac{3}{8}+1} -1 + 6 \ln 2 - \frac{\pi}{2}
\nonumber  \\
&&\hspace{0mm}
 + \frac{\pi}{\sqrt{2}} - \sqrt{2} \ln (\sqrt{2}+1) - 8 \sum_{i=1}^{[\sigma]+3} \frac{1}{8i-5}  \, ,
\label{eqB-14}
\end{eqnarray}
(v) for $\sigma=n+\frac{1}{2}$
\begin{eqnarray}
&&\hspace{-10mm}
C(\sigma) = \frac{2}{[\sigma]+\frac{1}{2}+1} -1 + 4 \ln 2 - 2 \sum_{i=1}^{[\sigma]+3} \frac{1}{2i-1}  \, ,
\label{eqB-15}
\end{eqnarray}
(vi) for $\sigma=n+\frac{5}{8}$
\begin{eqnarray}
&&\hspace{-10mm}
C(\sigma) = \frac{2}{[\sigma]+\frac{5}{8}+1} -1 + 6 \ln 2 + \frac{\pi}{2}
\nonumber  \\
&&\hspace{0mm}
 - \frac{\pi}{\sqrt{2}} - \sqrt{2} \ln (\sqrt{2}+1) - 8 \sum_{i=1}^{[\sigma]+3} \frac{1}{8i-3}  \, ,
\label{eqB-16}
\end{eqnarray}
(vii) for $\sigma=n+\frac{3}{4}$
\begin{eqnarray}
&&\hspace{-10mm}
C(\sigma) = \frac{2}{[\sigma]+\frac{3}{4}+1} -1 + 5 \ln 2 - \frac{\pi}{2}
\nonumber  \\
&&\hspace{0mm}
 - 4 \sum_{i=1}^{[\sigma]+3} \frac{1}{4i-1}  \, ,
\label{eqB-17}
\end{eqnarray}
(viii) for $\sigma=n+\frac{7}{8}$
\begin{eqnarray}
&&\hspace{-10mm}
C(\sigma) = \frac{2}{[\sigma]+\frac{7}{8}+1} -1 + 6 \ln 2 - \frac{\pi}{2}
\nonumber  \\
&&\hspace{0mm}
 - \frac{\pi}{\sqrt{2}} + \sqrt{2} \ln (\sqrt{2}+1) - 8 \sum_{i=1}^{[\sigma]+3} \frac{1}{8i-1}  \, ,
\label{eqB-18}
\end{eqnarray}
where $[\sigma]$ denotes the integer part of $\sigma$.  Finally, their numerical values are summarized in Table I.

\begin{table}[h]
\caption[]{Numerical values of $C(\sigma)$ for $\sigma=n+\alpha$. }
\begin{tabular}{crrrrrrr} \hline

     &  $n$=0 \, &  $n$=1 \, &  $n$=2 \, &  $n$=3 \, &  $n$=4 \, &  $n$=5 \, &  $n$=6 \, \\   \hline

 $\alpha$=0  &  0.8863  &  $-$0.4470  &  $-$1.030  &  $-$1.397  &  $-$1.664  &  $-$1.873  &  $-$2.046  \\

 $\alpha$=1/8  &  0.6159  &  $-$0.5407  &  $-$1.084  &  $-$1.435  &  $-$1.693  &  $-$1.897  &  $-$2.065  \\

 $\alpha$=1/4  &  0.3921  &  $-$0.6267  &  $-$1.136  &  $-$1.471  &  $-$1.720  &  $-$1.919  &  $-$2.085  \\

 $\alpha$=3/8  &  0.2026  &  $-$0.7061  &  $-$1.184  &  $-$1.506  &  $-$1.748  &  $-$1.942  &  $-$2.104  \\

 $\alpha$=1/2  &  0.0393  &  $-$0.7798  &  $-$1.231  &  $-$1.539  &  $-$1.774  &  $-$1.963  &  $-$2.122  \\

 $\alpha$=5/8  & $-$0.1038  &  $-$0.8485  &  $-$1.275  &  $-$1.572  &  $-$1.800  &  $-$1.985  &  $-$2.140  \\

 $\alpha$=3/4  & $-$0.2306  &  $-$0.9129  &  $-$1.317  &  $-$1.604  &  $-$1.825  &  $-$2.005  &  $-$2.158  \\

 $\alpha$=7/8  & $-$0.3443  &  $-$0.9733  &  $-$1.358  &  $-$1.634  &  $-$1.849  &  $-$2.026  &  $-$2.176  \\  \hline

\end{tabular}
\end{table}


\bibliography{apssamp}

\end{document}